\documentclass[aps,prl,twocolumn,groupedaddress,showpacs,%
amsmath,preprintnumbers]{revtex4}
\input{amssym}
\usepackage{hyperref}

\def\be {\begin{equation}}

\def\ee  {\end{equation}}

\def\bea {\begin{eqnarray}}

\def\eea {\end{eqnarray}}

\def\nn {\nonumber}
\begin{document}
\preprint{gr-qc/04}
\title{Background independent duals of the harmonic oscillator}
\author{Viqar Husain}
\email[]{vhusain@perimeterinstitute.ca} \affiliation{Perimeter
Institute for Theoretical Physics, 31 Caroline St. N, Waterloo ON,
Canada N2L 2Y5\\
Department of Mathematics and Statistics, University of New
Brunswick, Fredericton, NB, Canada E3B 5A3} 
\date{\today}
\begin{abstract}

We show that a class of topological field theories are quantum
duals of the harmonic oscillator.  This is demonstrated by
establishing a correspondence between the creation and
annihilation operators and non-local gauge invariant observables
of the topological field theory. The example is used to discuss
some issues concerning background independence and the relation of
vacuum energy to the problem of time in quantum gravity.

\end{abstract}

\maketitle

Two general themes have been a part of the dialogue in quantum
gravity in the past decade. These are the ideas associated with
the terms "background independence" and "duality."

The first of these stems from the intuition that a quantum theory
of gravity should be background independent. Exactly what
constitutes a background is a part of the debate in quantum
gravity \cite{isham,lee}, since there are levels of structure,
from a set of points to a manifold with a metric, that may be
fixed when formulating a classical or quantum theory.  What is
clear is the semi-classical requirement that in weak gravity an
effective metric should emerge for some quantum states, and that
in strong gravity no metric should be preferred. For example, the
formal statement of the partition function for quantum gravity as
a functional integral over all Lorenzian four-geometries
illustrates 4-metric independence, up to possible fixed structure
at boundaries. In the canonical framework, the background
structure includes fixing the 4-manifold to be $R\times \Sigma$,
where space $\Sigma$ also has fixed topology. The same structure
would exist in the path integral framework in a fixed time gauge.

An operating definition of background independence may be taken to
be metric independence. Classical general relativity is background
independent in the sense that the metric is varied in the action,
and therefore does not constitute a fixed structure. In contrast
the definition of graviton is background dependent since it is a
perturbation on a fixed metric. It is widely accepted that a
non-perturbative theory of quantum gravity should not be one that
prefers a fixed background metric at the fundamental level.

Background independence is  not well understood for several
reasons. The most prominent among these is that no concrete
examples of background independent quantum field theories are
known in which a metric is emergent in the semi-classical limit,
where one expects to obtain the approximation of quantum fields on
a fixed background. The issue is compounded by the fact that no
background independent formulation is known even for the free
scalar field that can make contact with conventional quantum field
theory dynamics where the propagator is fundamentally metric
dependent \cite{carlo}.

Duality on the other hand appears to be better understood, at
least in some contexts. In one of its incarnations, it is the idea
that two apparently distinct theories, not necessarily in the same
spacetime dimension, are equivalent at the quantum level.
"Equivalent" means that there is a precise correspondence between
operators and quantum states in the dual theories, a relation
between their coupling constants, and a matching of the spectra of
operators, at least in some limits.

The earliest example of a duality dates to the seventies, when the
first equivalence between field theories in 2-dimensions was
established. This is the duality of the massive Thirring  and
sine-Gordon theories \cite{coleman}. More recent examples of
dualities  are the series of AdS/CFT correspondences, and mirror
symmetry. The former is a proposed duality between certain
supersymmetric conformal field theories on Minkowski spacetime and
string theory on asymptotically anti-deSitter spaces
\cite{adsrev}. The mirror symmetries are equivalences between the
quantum theories of certain 2-dimensional supersymmetric sigma
models whose the target spaces are Calabi-Yau manifolds
\cite{mirror}.

Duality and background independence become related if one of a
dual pair of theories is a theory of geometry, and the other is a
conventional theory describing field dynamics on a fixed metric
background. Dualities of this type could perhaps allow a
background independent question, such as a quantum gravity
transition amplitude, or questions about black hole physics, to be
reformulated and answered in a conventional fixed-metric setting.
The AdS/CFT correspondences are often cited as providing an
example of this sort. So in principle at least, they provide a
scenario where quantum gravity questions are addressable in the
CFT, and vice versa. This has motivated discussion on whether {\it
any} CFT has a geometric, or background independent dual theory
\cite{strings05}.

It is therefore of interest to seek other examples of dualities of
this type and to see what can be learned from them. It is also
important to distinguish the cases where there is imposed
asymptotic background structure, as in the asymptotically flat or
anti-deSitter cases, from the cases where space has no boundary.
This is because a gravity theory with no classically fixed time
has a hamiltonian constraint rather than a usual hamiltonian, so
any comparison of its operator spectra with those of a proposed
dual theory with a fixed time concept requires interpretive care.

Motivated in part by this debate, we point out in this note an
exact duality between a topological (and therefore background
independent) field theory in $n$ dimensions $n>2$, and the simple
harmonic oscillator in 0+1 dimensions. This is done by finding the
fully gauge invariant observables of the topological theory, and
showing that certain functions of these satisfy the algebra of the
creation and annihilation operators. Our aim is to clarify using
this example the extent to which quantum gravity questions may be
addressable via a dual background dependent theory.

 The topological field theories of interest here are
the so called BF models on an n-dimensional manifold $M$
\cite{hor,vhtqm,blau}. The action is
\be S = k \int_M {\rm Tr}[B\wedge F(A)]. \ee
where $A$ is a 1-form, $F(A)$ is its curvature, and $B$ is an
$n-2$ form. The trace is in the Lie algebra in which the fields
are valued.

Our main point concerning duality is illustrated by the Abelian
theory in four dimensions on a manifold $M \sim \Sigma \times R$,
where $\Sigma$ is a 3-manifold without boundary. Since the action
is first order, it is easy to put into canonical from:
\be S = k\ \int_{\Sigma \times R} 2\epsilon^{0abc} \left[
B_{ab}\partial_0A_c + B_{0a}F_{bc} - B_{ab}\partial_c A_0 \right],
\ee
where $a,b,c$ are indices in $\Sigma$. The canonical phase space
coordinates are therefore $(A_a,E^a)$, where $E^a =
\epsilon^{0abc} B_{bc}$, which satisfy the Poisson bracket
relations
\be
\{A_a(x,t),E^{b}(y,t)\}= \frac{1}{k}\  \delta_a^b\delta^3(x,y)
\ee
The Hamiltonian is a linear combination of the constraints
\be F_{ab} = 0, \ \ \ \ \ \ \ \ \ \partial_aE^a = 0, \ee
obtained by varying the action with respect to $B_{0a}$ and $A_0$.

Since the constraints generate gauge transformations of the
canonical variables via Poisson brackets, the gauge invariant
observables ${\cal O}(E,A)$ are defined by the Poisson bracket
conditions
\be \left\{ {\cal O}(E,A), C(E,A) \right\} = 0, \ee
where $C$ denotes the two constraints.

In the present case observables satisfying this condition are the
non-local functionals
\bea
{\cal O}_1(A,\gamma) &=& \int_\gamma ds\ \dot{\gamma}^a A_a, \\
{\cal O}_2(E,S) &=& \int_S d^2\sigma\ n_aE^a. \label{obs}
\eea
These are parametrized by embedded loops $\gamma$ and surfaces $S$
in $\Sigma$, and $n_a$ is a one form field defining the surface
$S$ ($\epsilon^{0abc}n_c$ is the area 2-form and $\dot{\gamma}^a$
is tangent vector to the loop $\gamma$). These observables satisfy
the Poisson algebra
\bea
 \left\{ {\cal O}_1(A,\gamma), {\cal O}_1(A,\beta)\right\} &=& 0,\\
\left\{ {\cal O}_2(E,S), {\cal O}_2(E,S') \right\} &=& 0,\\
 \left\{ {\cal O}_1(A,\gamma), {\cal O}_2(E,S) \right\}
 &=& \frac{1}{k}\ c(\gamma, S) C_1(A,\gamma)
\eea
where
\be c(\gamma,S) = \int ds\int d^2\sigma\
\dot{\gamma}^an_a\delta^3(\gamma(s) - S(\sigma)), \ee
counts the intersections of the loop with the surface. The last
Poisson bracket vanishes if $\dot{\gamma}^an_a = 0$, or if the
loop and surface have no points of intersection.

As is well known, first class constraints have two properties.
They generate gauge transformations, and they restrict the
dynamics to the surface in the phase space defined by their strong
imposition. Off the constraint surface there are an uncountable
infinity of observables, because of the number of possible loops
and surfaces in $\Sigma$.  On the constraint surface however, most
of these vanish because of the flat connection constraint; only a
finite number that depend on the non- contractible loops and
surfaces in $\Sigma$ remain. These capture topological information
about $\Sigma$. (From a covariant point of view this is seen by
the fact that the equations of motion $dB=0$ and $F=dA=0$ have a
finite dimensional solution space given by the dimensions of the
cohomology groups of $M$.)

To proceed further we must fix the topology of $\Sigma$. This
determines the number of independent observables, and hence the
number of degrees of freedom. Perhaps the simplest example is
provided by the case $\Sigma \sim S^1\times S^2$ for which there
is one non-contractible loop and surface, with $c(\gamma,S)=1$.
Thus there are exactly two degrees of freedom, which satisfy the
Poisson algebra
\be
 \left\{ C_1, C_2\right\} = \frac{1}{k}.
\ee

A quantization  for this spatial topology is obtained by realizing
this as a commutator on a Hilbert space. An occupation number
representation is obtained by defining the operators
\be \hat{a}^\pm = \sqrt{\frac{k}{2\hbar}}\ \left(\hat{C}_1 \pm i
\hat{C}_2 \right) \ee
with their usual action. This establishes a duality with the usual
algebraic quantization of the harmonic oscillator, up to the issue
of Hamiltonian. This is discussed below.

Other spatial topologies in the abelian theory give more
observables in the reduced theory. For example the case $\Sigma
\sim T^3$ has three pairs of surface and loop observables, which
is equivalent to three uncoupled oscillators.

A similar correspondence with the harmonic oscillator exists for
the non-abelian theory. The (unreduced) phase space of the theory
is that of Yang-Mills theory, but with additional constraint
functions. These are
\be D_aE^{ai} =0,\ \ \ \ F_{ab}^i=0, \ee
where $i$ is a Lie algebra index. The observables for the theory
are a bit more involved than for the Abelian theory \cite{vhtqm}.
As for any theory with a Gauss law, these are made from the
holonomy of the connection $A_a^i$ around loops $\alpha$
\be U_\alpha(A) = {\rm P\ exp}\int_\alpha ds\ A_a(\alpha(s))\
\dot{\alpha}^a(s). \ee
The first type of observable is the trace of holonomy
\be
T^0(A;\alpha)={\rm Tr}\ [U_\gamma(A)].
\ee
The second type depends on a loop $\alpha$ and a closed 2-surface
$S$,
\be
T^1(A,E;\alpha,S) = \int_S d^2\sigma\ n_a{\rm Tr}
\left[E^a(\sigma)U_\alpha(\sigma)\right].
\ee
The integrand in the latter is a function of the holonomy of loops
$\alpha$ whose base point lies in the surface $S$. The surface
integral is over all locations of the base point in $S$. Both
observables may be constructed in a fixed representation of the
group, which we take to be the fundamental one. It is readily
verified that $T^1$ Poisson commutes with the constraints, and so
is a fully gauge invariant observable. $T^0$ satisfies this
trivially. ($T^1$ is an integrated version of one of a series of
partially gauge invariant observables for quantum gravity
introduced in Ref. \cite{rs}.)

The Poisson algebra of the observables for the $SU(2)$ theory has
a rather nice structure:
\be \{ T^0(\alpha),T^0(\beta) \} = 0, \ee
\bea\{ T^0(\alpha),T^1(S,\beta)\} &=& ic(\alpha,S)\nn \\
 &&\times \left[T^0(\alpha\circ\beta) - T^0(\alpha\circ\beta^{-1})\right],
 \eea
\bea &&\{T^1(\alpha,S), T^1(\beta,S'\} = \nn\\
&& ic(\alpha,S')\left[T^1(\alpha\circ\beta,S)-T^1(\alpha\circ\beta^{-1} \right] \nn\\
&&-ic(\beta,S)\left[T^1(\beta\circ\alpha,S)-T^1(\beta\circ\alpha^{-1}\right],
 \eea
where $\alpha\circ\beta$ etc. denote a product of holonomies for
the respective loops.

Let us again consider the case $\Sigma\sim S^1\times S^2$. The
non-trivial observables on the reduced phase space are again
specified by the non-contractible loops and surfaces in $\Sigma$:
all the loops and surfaces that wrap once around the circle and
the sphere respectively give equivalent non-trivial observables.
Thus there is exactly one basic observable of each type, which we
denote by $T^0(a)$ and $T^1(a,S)$, where $(a,S)$ denote the circle
and the sphere. The observable algebra on the reduced phase space
simplifies to
\bea
\{T^0(a), T^0(a)\} &=& 0=\{T^1(a,S), T^1(a,S)\}, \nn \\
\{T^0(a),T^1(a,S)\} &=& i\left[(T^0(a^2) - T^0(aa^{-1})\right].
 \eea
The last equation may be rewritten using the trace identity ${\rm
Tr}(A){\rm Tr}(B) = {\rm Tr}(AB)+{\rm Tr}(AB^{-1})$ for $SU(2)$
matrices $A,B$, and the fact that $T^0(aa^{-1})$ is the trace of
the $2\times 2$ identity matrix (since we are using the
fundamental representation of SU(2)). This gives
\be \{T^0(a),T^1(a,S)\} = i [T^0(a)^2 - 4]. \ee

A correspondence of this algebra with that of the harmonic
oscillator is established by defining the creation and
annihilation variables by
\be A^\pm = \frac{1}{\sqrt{2}}\ \left[T^0 \pm \frac{T^1}{(T^0)^2-
4}\right]. \ee

It is possible to proceed similarly with other spatial topologies
and gauge groups. The common feature in all examples is the
expression of one or more copies of the oscillator variables as
functions of the basic gauge invariant non-local observables of
the topological field theory. The central and non-trivial property
that permits this is that the observable algebra of the BF
theories we have discussed not only closes, but is also simple
enough to reveal the combinations of functions that are
canonically conjugate.

Although of mathematical interest, the central question concerning
dualities between background independent theories and conventional
ones is what can be learned about the physics of one theory from
the other.  For the cases discussed here, it is fair to ask what
one can learn about the harmonic oscillator from the topological
field theory, beyond the identification of oscillator variables.

A first observation concerns the problem of time \cite{karel}. The
topological examples provided here are {\it fully} background
indepedendent in the sense that there are no fixed asymptotic
structures. This is unlike the AdS/CFT correspondences, where from
the start the asymptotic isometry group of the spacetimes under
consideration are prescribed, and carry with them a notion of
time. The important physical difference between topological field
theories and the harmonic oscillator is that the latter has a
non-vanishing Hamiltonian which gives non-trivial dynamics with
respect to an external (Newtonian) time. On the other hand, the
Hamiltonian of any theory with time reparametrisation invariance
is a phase space constraint so that time evolution is pure gauge,
and gauge invariant observables are also constants of the motion.
Therefore the Heisenberg evolution equations are trivially
satisfied for the topological field theory observables. This means
that one cannot "see" the dynamics of the oscillator in the
topological theory, although the "number operator" of the latter
is the dual of the oscillator Hamiltonian:
\be H_{osc} \leftrightarrow (a^\dagger a)_{tft}
\label{aan}
\ee

In the cases where there is no classically fixed asymptotics,
there are two known ways to introduce a time variable in a
background independent theory: a classical time gauge fixing, or a
relational time where a "slow" phase space variable  is chosen as
an internal clock, and evolution of other phase space variables
are viewed with respect to it.

Without such choices, perhaps the best that can be done is a {\it
kinematic} correspondence like the one presented here, where an
equivalence is established between observable algebras up to time
evolution on one side: the oscillator algebra of the operators
$\hat{a}(t),\hat{a}^+(t)$ at any value of Newtonian time maps to
the timeless algebra of the gauge invariant observables of the
topological theory.   Equivalently one can say that the
correspondence is at the level of the time independent
Schr{\"o}dinger equations of the two theories, where on the
topological theory side this equation is $(a^\dagger a)_{tft}
|\psi\rangle = E |\psi\rangle$. There is no non-trivial evolution
equation for the fully gauge invariant variables, since these are
(by definition) also constants of the motion.

The second observation concerns the cosmological constant, and
arises from the issue of the matching of spectra. With the
correspondence of Eqn. (\ref{aan}), the eigenvalues differ by a
constant shift -- the ground state energy of the oscillator. The
number operator of the topological theory does not have an
"energy" interpretation beyond that suggested by the
correspondence. However it does raise an issue which is ultimately
connected with the cosmological constant (or vacuum energy)
problem.

The conventional vacuum energy problem arises in the context of
quantum fields on a fixed background spacetime \cite{revs}. The
association of matter vacuum energy with the cosmological constant
is made using the semiclassical equation
\be
G_{ab} + \Lambda^{(f)} g_{ab}=  8\pi G \langle \hat{T}_{ab}
\rangle
\ee
where $\Lambda^{(f)}$ is a fundamental (or bare) cosmological
constant. From this equation the predicted cosmological constant
is given by
\be \Lambda^{(theory)} = \Lambda^{(f)} - 2\pi G\ \langle
\hat{T}_{ab}\rangle\ g^{ab}, \ee
where the expectation value is taken in some "vacuum" state. If
the background metric has a timelike Killing vector field, there
is a preferred matter Hamiltonian, and hence a vacuum. For
dynamical metrics on the other hand, the additional assumption of
a time gauge choice is needed to identify a Hamiltonian and its
vacuum. Thus in general it is evident that $\Lambda^{(theory)}$
{\it is dependent on the choice of time}. With an assumed Planck
scale cutoff and $\Lambda^{(f)}=0$, the statement of the
cosmological constant problem is the oft quoted discrepancy
\cite{revs} $\Lambda^{(theory)}/\Lambda^{(obs)} = 10^{120}$. If
$\Lambda^{f}\ne 0$, this translates to a "fine tuning" problem.

To reformulate all this at a more fundamental level one needs a
notion of time, and its associated non-vanishing quantum gravity
Hamiltonian density $\hat{H}(\hat{q},\hat{\pi};
\hat{\phi},\hat{P}_\phi; \Lambda^{(f)},g_i;t)$. This is a function
of the gravity $(q_{ab},\pi^{ab})$ and matter $(\phi,P_\phi)$
operators, the cosmological constant, and other coupling constants
$g_i$. Furthermore, it must have  {\it explicit} dependence on a
time variable $t$, however it arises from a fundamental background
independent theory. (This is evident for example in the reduced
Hamiltonians obtained by imposing various time gauge fixings in
cosmological models.)

The task is to find ground state(s) $|q,\phi\rangle_0$, of this
Hamiltonian and compute the vacuum energy.  It is at this stage
that there may be an emergent "cosmological constant problem" if
the energy of the relevant state of $\hat{H}$ does not match the
observed one, ie. if it turns out that
\be
\ _0\langle q,\phi|\ \hat{H}  |q,\phi\rangle_0 \equiv
\rho_0(\Lambda^{(f)},g_i;t) \sim \rho^{(obs)}
\ee
requires fine tuning of $\Lambda^{(f)}$ and $g_i$ when the present
value of time is inserted on the left hand side of this equation.
Furthermore, since the expectation value has explicit time
dependence, it is evident that to agree with observations, the
observed value of vacuum energy density must not be a fixed
constant.

What is apparent from these observations is that if one starts
from a background independent gravity-matter theory,  the problem
of time must be solved before one can even ask if there is a
cosmological constant problem.

In summary, what we have learned from  the duality example given
here is that, although it may be possible to establish exact
dualities between a background independent theory and a background
dependent one by presenting an operator dictionary, the challenge
remains to establish a dynamical correspondence between physical
processes. Both this, and a more fundamental quantum gravity based
statement of the cosmological constant problem, require a solution
of the problem of time in quantum gravity.
\medskip

 \acknowledgments{This work was supported in part by the Natural
Science and Engineering Research Council of Canada. I would like
to thank Laurent Freidel, Stefan Hofmann, Lee Smolin and Oliver
Winkler for discussions.}

\end{document}